 \documentclass[final,3p,times,twocolumn, authoryear]{elsarticle}

\usepackage{graphicx}
\usepackage{booktabs}
\usepackage{siunitx}
\usepackage{amsmath}
\usepackage{lineno}
\usepackage{float} 
\usepackage{natbib}
\usepackage{color}
\usepackage[colorlinks=true, allcolors=blue]{hyperref}
\usepackage[capitalize]{cleveref}
\usepackage{caption}


\begin{document}

\begin{frontmatter}



\title{Influence of dissolved gas concentration on the lifetime of surface bubbles in volatile liquids} %% Article title


\author[label1]{Xin Li}
\author[label1,label2]{Yanshen Li\corref{cor1}}
\cortext[cor1]{Email address for correspondence: \href{mailto:liyanshen@ucas.ac.cn}{liyanshen@ucas.ac.cn}}

\affiliation[label1]{organization={School of Engineering Science, University of Chinese Academy of Sciences},
        %addressline={No. 380 Huaibeizhuang, Huairou District},
        city={Beijing},
        postcode={101408},
        country={PR China}}

\affiliation[label2]{organization={State Key Laboratory of Nonlinear Mechanics, Institute of Mechanics, Chinese Academy of Sciences},
        %addressline={No. 380 Huaibeizhuang, Huairou District},
        city={Beijing},
        postcode={100190},
        country={PR China}}

%% Abstract
\begin{abstract}
Bubbles at the air--liquid interface are important for many natural and industrial processes. Factors influencing the lifetime of such surface bubbles have been investigated extensively, yet the impact of dissolved gas concentration remains unexplored.Here we investigate how the lifetime of surface bubbles in volatile liquids depends on the dissolved gas concentration. The bubble lifetime is found to decrease with the dissolved gas concentration. Larger microbubbles at increased gas concentration are found to trigger bubble bursting at earlier times. Combined with the thinning rate of the bubble cap thickness, a scaling law of the bubble lifetime is developed. The scaling is also found to be independent of factors like container type, liquid pool depth and bubble size. Our findings may provide new insight on surface bubble lifetime and foam stability. 
\end{abstract}

%\item Dissolved gas concentration are found to decrease the lifetime of surface bubbles
%\item Larger mircobubbles at increased gas concentration trigger the earlier burst of surface bubbles
%\item For volatile liquids, the relationship between bubble lifetime $t_l$ and dissolved gas concentration $c$ follows $t_l\sim c^{-3/2}$ 

%% Keywords

\end{frontmatter}



\section{Introduction}

Bubbles are omnipresent in nature and industrial applications, such as in floatation \citep{peleka2018perspective}, renewable energy production \citep{ardo2018pathways, chatenet2022water}, inkjet printing \citep{lohse2022fundamental}, thermal management \citep{mohanty2017critial, yang2023review}, ultrasound diagnostics \citep{dollet2019bubble}, etc. For comprehensive reviews on bubbles, we refer to \citet{lohse2018bubble} and \citet{garbin2025bubbles}.

Among them, bubbles at the air/liquid interface, also called surface bubbles, are important for climate change, food and household industry. For example, when bubbles on the sea surface burst, they release aerosols into the atmosphere \citep{lhuissier2012bursting, villermaux2022bubbles, jiang2024abyss}, which influences the air-sea mass exchange \citep{deike2022mass}. In carbonated beverages, surface bubbles may form a layer of foam, whose quality is crucial for the flavor and visual appeal of the beverage \citep{viejo2019bubbles}. In bath and cleaning products, surface bubbles amplify detergent-dirt contact, boosting cleaning efficiency and user comfort \citep{jin2022environment}. The stability and lifetime of surface bubbles are found to be vital to these applications. Thus, extensive studies have investigated factors that influence the lifetime of surface bubbles \citep{miguet2021life, poulain2018ageing,shaw2024film,auregan2025drainage}. Surfactants were found to increase the lifespan of surface bubbles by reducing surface tension and inducing Marangoni flows, which hinders film drainage due to gravity \citep{sonin1993role, adami2015surface}. Temperature gradients or concentration gradients, caused by evaporation \citep{menesses2019surfactant, lorenceau2020lifetime} or imposed directly \citep{nath2022thermal}, could also increase the lifetime by introducing Marangoni flows on the bubble cap, which compensates for the film drainage. could also increase the lifetime by introducing Marangoni flows on the bubble cap,
which compensates for the film drainage. Viscosity also increases the lifetime by directly slowing the film drainage \citep{debregeas1998life, oratis2020new}. Salinity was found to influence the bubble lifetime in a non-monotonic way \citep{magdalena2017effects}. However, among all the factors that influence the surface bubble lifetime, the impact of dissolved gas concentration has not been explored, even though it is strongly relevant in the food industry. 

We investigate the influence of dissolved gas concentration on the lifetime of surface bubbles in volatile liquids. The dissolved gas concentrations are varied by depressurizing or pressurizing the liquid. Dissolved oxygen concentrations are measured to calculate the gas oversaturation/undersaturation. Surface bubbles are generated by injecting a certain volume of air into the bulk liquid and letting it rise. Lifetime of surface bubbles are found to decrease while increasing the dissolved gas concentration. After analyzing the recorded bursting events, we find that microbubbles formed in the bulk liquid are responsible for the earlier bursting: When large enough microbubbles enter the bubble cap, they rupture the film by inducing an additional local thinning of the film. We also find that the average microbubble diameter increases almost linearly with the dissolved gas concentration. Combined with the thinning rate of the film thickness, a scaling law is developed to explain the dependence of surface bubble lifetime on the dissolved gas concentration. Other factors like container type, liquid pool depth and bubble size are varied, which are found to only change the absolute value of bubble lifetime, but not influencing the scaling law.

\section{Experimental procedures and methods}
\label{ExpSetup}
\begin{figure}[H]
    \centering
    \includegraphics[width=0.5\textwidth]{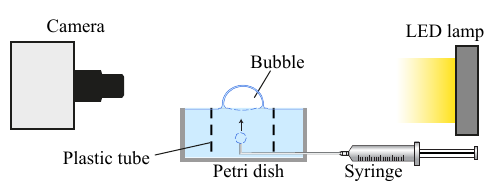}
    \caption{The experimental setup to investigate the dependence of surface bubble lifetime on factors such as dissolved gas concentration, container type, pool depth, etc. Bubbles are generated by injecting air at a constant flow rate through a \SI{0.5}{mm} inner diameter needle below the liquid surface. A plastic tube shorter than the liquid depth is put in the liquid pool to fix the position of surface bubbles.}
    \label{fig:setup}
\end{figure}

At room temperature, a petri dish (STEENA, China) of diameter \SI{3.5}{cm} and depth \SI{10}{mm} was filled with liquids of different dissolved gas concentrations. Isopropanol (Mreda,purity \SI{99.5}{\%}) was used as the liquid. The petri dish was rinsed with ethanol and deionized water (Hhitech, Master Touch-RUV, China) then dried with compressed air before use. As shown in Fig. \ref{fig:setup}, a needle of inner diameter $\SI{0.51}{mm}$ was immersed in the liquid, which was connected to a syringe mounted on a syringe pump (Leadfluid, TFD02, China). An air bubble of volume \SI{90}{\micro\litre} was injected at a flow rate \SI{42.75}{mL/min} so as to form a hemispherical bubble at the surface. A  plastic tube shorter than the liquid depth was vertically placed at the center of the petri dish to prevent the surface bubble from moving around.Both the petri dish and the plastic tube are made of polyethylene (PE). A collimated LED (Thorlabs, LEDD1B, USA) and a camera (Nikon D850, at \SI{60}{fps} and $1920\times 1080$ resolution) connected with a long working distance zoom lens system (Thorlabs, MVL12X12Z plus 2X lens attachment, USA) were used to record the life time $t_l$ of the bubble. To find out the influence of container type, pool depth and bubble size on bubble lifetime, the experiments were repeated by changing the container to cubic glass containers made of quartz, increasing the pool depth to \SI{20}{mm} and \SI{40}{mm}, and changing the bubble volume from \SI{90}{\micro\liter} to \SI{25}{\micro\liter} and \SI{55}{\micro\liter}, see Table \ref{tab:parameters} for the detailed parameters of each experimental condition.

To control the dissolved gas concentration, different methods were used to prepare the undersaturated and oversaturated liquids. For undersaturation, around \SI{100}{mL} liquid was put in a vacuum-desiccator at around \SI{1000}{Pa} for 5 to \SI{30}{min} to control its gas concentration. For oversaturation, around \SI{100}{mL} liquid was pressurized with air at 1.5 to \SI{4.5}{bar} for 2 to \SI{24}{h}. Then the undersaturated or oversaturated liquid was poured into a wide-mouth bottle to measure its dissolved oxygen concentration $c_\mathrm{o}$ by an optical dissolved oxygen meter (Mettler-Toledo, S900-K, USA). It takes about \SI{30}{s} to measure the dissolved oxygen concentration $c_\mathrm{o}$ in the liquid. Right after the $c_\mathrm{o}$ had been measured, the liquid was filled into the container to start the experiment and the bottle was sealed to prevent the gas concentration in it from changing. To minimize the change of gas concentration in the liquid in the container due to gas exchange with air during the experiments, the undersaturated (oversaturated) liquid was used for no longer than \SI{5}{min} (\SI{2}{min}) unless the bubble lifetime exceeds this value. After this time (or after bubble rupture), liquid in the sealed bottle was taken again to fill the container to continue the experiment. These two time intervals were selected because it was found that within these time intervals, the gas concentration in the liquid does not change much (see Supplementary Material for more details).

To capture the detailed rupture processes of the bubbles, a high speed camera (Photron, Nova S12, Japan) at frame rates ranging from 3000 to \SI{25000}{fps} was used. Liquid film thickness $h$ of the bubble was calculated using the Taylor-Culick method \citep{taylor1956fluid,2020Experimental}:
\begin{equation}
h = \frac{2 \sigma}{\rho v^2},
\end{equation}
where $\sigma$ is the surface tension and $\rho$ is the density of the liquid, $v$ is the retraction speed of the edge of the hole in the film, which is measured by the recorded high speed video. The initial film thickness $h_i$ of the bubble was also measured using this method by puncturing the bubble with a needle immediately after it is generated. The density and surface tension of isopropanol at \SI{25}{\degreeCelsius} and \SI{1}{atm} are \SI{0.781}{g/cm^3} and \SI{20.4}{mN/m} \citep{haynes2016crc}, respectively. 

\section{Results and discussions}

In all the experiments, the pool liquid is exposed to air, even during depressurization and pressurization, so that the pool liquid is undersaturated or oversaturated with gas components in the air. Let $c$ denote the gas concentration in the pool liquid and $c_\mathrm{sat}$ the saturated gas concentration, we obtain the gas oversaturation $\zeta$ which is defined as
\begin{equation}
\zeta=\frac{c-c_\mathrm{sat}}{c_\mathrm{sat}}.
\end{equation}
Naturally, the liquid is oversaturated (undersaturated) when $\zeta>0$ ($\zeta<0$). Because different gas components in the air dissolve in or get released from the liquid at the same ratio (see \ref{app:A}), i.e., $c_\mathrm{o}/c=\beta$ where $\beta$ is a constant independent of pressure, we have 
\begin{equation}
\zeta=\frac{c-c_\mathrm{sat}}{c_\mathrm{sat}}=\frac{c_\mathrm{o}-c_\mathrm{o, sat}}{c_\mathrm{o, sat}},
\end{equation}
where $c_\mathrm{o, sat}$ is the saturated oxygen concentration in the pool liquid, which is measured to be \SI{8.3}{mg/L} at room temperature (\SI{23}{\degreeCelsius}) and \SI{1}{atm}. 
\color{black}

\begin{figure}[H]
    \centering
    \includegraphics[width=0.5\textwidth]{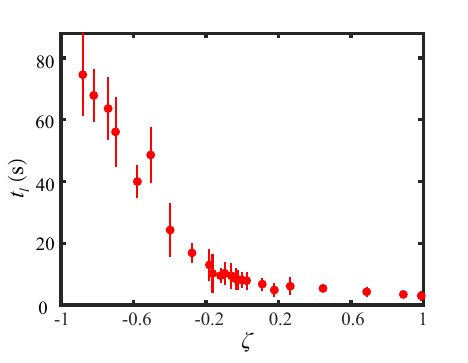}
    \caption{Isopropanol bubble lifetime $t_l$ at different dissolved gas concentrations represented by the gas oversaturation $\zeta$. Each data point is the average of at least 13 experimental repetitions and the errorbars are the standard deviation}.
    \label{fig:lifetime}
\end{figure}

Then, the influence of dissolved gas concentration, represented by the gas oversaturation $\zeta$, on the bubble lifetime $t_l$ is shown in Fig. \ref{fig:lifetime}. The data are collected from 363 experiments. It is clearly shown that the surface bubble lifetime $t_l$ decreases as the dissolved gas concentration in the liquid increases. Especially, the bubble lifetime in the undersaturated liquid decreases sharply as the gas concentration increases. As the dissolved gas becomes saturated , the bubble lifetime starts to decrease slowly. This trend continues even when the liquid is oversaturated. 

\begin{figure}[b]
    \centering
    \includegraphics[width=0.5\textwidth]{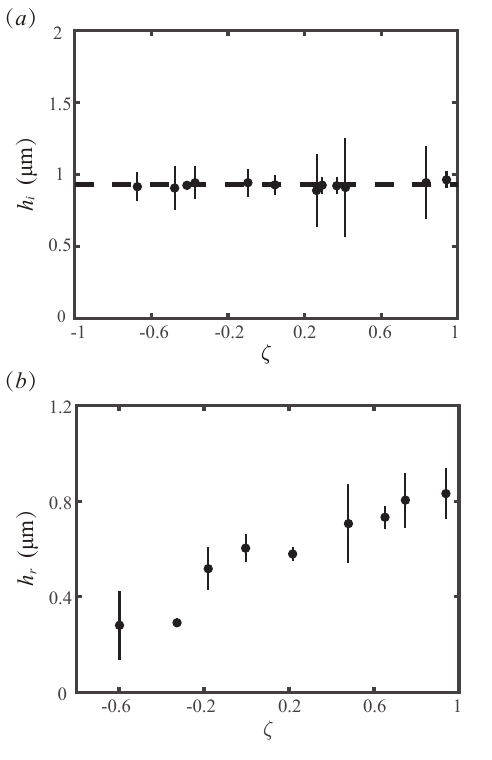}
    \caption{($a$) Initial film thickness $h_i$ of isopropanol bubbles at different dissolved gas concentrations represented by the gas oversaturation $\zeta$. The initial film thickness $h_i$ is measured by puncturing the bubble with a needle 0.8-1.3 seconds after it is generated. Each point is the average of 15 bubbles and the errorbars are the standard deviation. The horizontal dashed line is the overall mean $h_i\approx\SI{0.93}{\micro\meter}$. ($b$) Film thickness at spontaneous rupture $h_r$ for different gas concentrations.}
    \label{fig:initial_thickness}
\end{figure}

The bubble cap is made of a thin layer of liquid, whose stability and rupture determines the lifetime. The rupture of surface bubbles can be divided into two stages \citep{poulain2018ageing}: (i) Deterministic thinning of the liquid film due to gravity. (ii) Rupture of the film due to internal or external disturbances, such as particles or bubbles, which rupture the film by nucleating a hole large enough in the liquid film. Of all the factors that influence the thinning rate of the liquid film, such as surface tension and liquid evaporation \citep{menesses2019surfactant, lorenceau2020lifetime, nath2022thermal, debregeas1998life, oratis2020new}, dissolved gas concentration does not seem to influence any of them. Especially, a thin liquid layer on the pool will become gas-equilibrated with the atmosphere immediately after it is exposed to air, thus the surface bubble is almost always air saturated. Therefore, we speculate that the increased gas concentration in the liquid does not influence the film thinning process, but affects the surface bubble lifetime by bringing strong enough disturbances earlier to the film, consequently leading to its earlier rupture. We will see later that these ``earlier disturbances'' are microbubbles formed in the liquid pool which later rises to the bubble cap and triggers the rupture. This speculation is supported by measuring the initial film thickness $h_i$ of the isopropanol bubbles and the film thickness at rupture $h_r$. The film thicknesses are measured by Taylor-Culick method, see Section \ref{ExpSetup}. The initial film thickness $h_i$ is measured by puncturing the bubble with a needle 0.8-1.3 seconds after it is generated and $h_r$ is measured at the moment of spontaneous rupture. Fig. \ref{fig:initial_thickness}($a$) shows the initial film thickness $h_i$ of the surface bubbles at different dissolved gas concentrations.  It is clear that the initial film thickness of the bubbles does not vary with the dissolved gas concentration, and the average value is found to be $h_i\approx\SI{0.93}{\micro\meter}$. Fig. \ref{fig:initial_thickness}($b$) shows the film thickness $h_r$ at rupture. It can be seen that as the dissolved gas concentration increases, the film thickness at rupture $h_r$ also increases. This suggests that by increasing the dissolved gas concentration, the decrease in bubble lifetime is due to an early rupture of the film. Notice that the largest film thickness at rupture is $h_r=\SI{0.8}{\micro\meter}$, smaller than the initial film thickness $\SI{0.93}{\micro\meter}$.

\begin{figure*}[t]
    \centering
    \includegraphics[width=\textwidth]{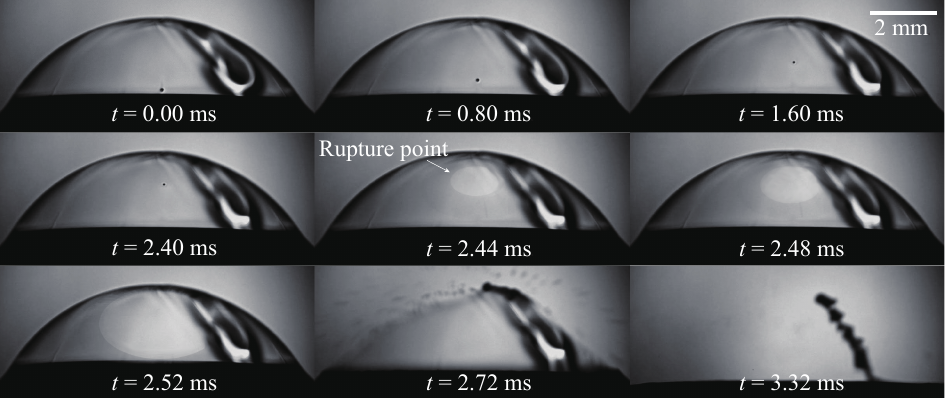}
    \caption{Snapshots of a rupture event of an isopropanol bubble. The liquid is oversaturated and its dissolved oxygen concentration is $c_\mathrm{o}= \SI{9.33}{mg/L}$. A small particle (the black dot) rises from the bottom to the bubble cap (from $t=\SI{0}{ms}$ to \SI{0.24}{ms}), immediately after (\SI{0.04}{ms} later), a hole appears and expands at the position of the particle, which later leads to the rupture of the bubble. The scale bar is \SI{2}{mm}.}
    \label{fig:microbubble_rupture}
\end{figure*}

\begin{figure}[b]
    \centering
    \includegraphics[width=0.5\textwidth]{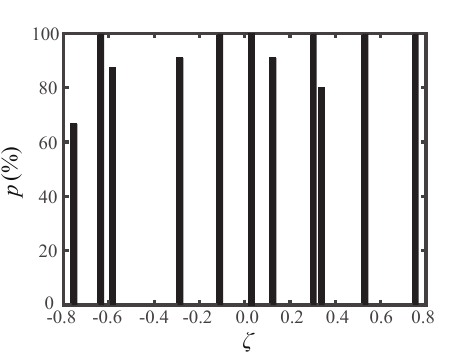}
    \caption{Probability of particle-triggered-bursting of isopropanol surface bubbles at different dissolved gas concentrations represented by the oversaturation $\zeta$. 3 to 11 bursting events were counted at each oversaturation, and in total 64 out of 69 bursting events were found to be triggered by a particle, which gives an overall probability of \SI{92.8}{\%}.}
    \label{fig:rupture_histogram}
\end{figure}

We then observe the bubble closely to see what triggers the rupture of the film. It was found that a small particle rising in the film is responsible for the initiation of the hole and the subsequent rupture, see Fig. \ref{fig:microbubble_rupture} for a typical rupture event recorded by the high speed camera (see also Movie S1). As can be seen, a small particle (the black dot) first rises in the liquid film from the bottom of the bubble (from $t=\SI{0}{ms}$ to \SI{0.24}{ms}). Immediately after (\SI{0.04}{ms} later), a hole appears at the position of the particle. This hole expands due to surface tension, which finally leads to the rupture of the film. These snapshots show that this single rupture event was triggered by internal disturbances induced by the particle, which induces an additional local thinning of the film \citep{poulain2018ageing}. In order to find out the probability of bubble bursts due to the same reason, i.e., caused by a particle, more bursting events at different dissolved gas concentrations are recorded. Those with a clear evidence that the rupture was triggered by a particle (similar to that show in Fig. \ref{fig:microbubble_rupture}) were counted, and the probability is shown in Fig. \ref{fig:rupture_histogram}. It is found that 64 out of 69 bursting events are confirmed to be triggered by a particle, giving an overall probability of \SI{92.8}{\%}. The real probability of particle-triggered-bursting should be larger than this value, because in some cases the particle is not visible to the camera and thus not counted: For example, when the particle trajectory happens to be overlapping with the bubble's outline, it cannot be recorded by the camera.

Nevertheless, the probability of particle-triggered-bursting being larger than \SI{92.8}{\%} further confirms our speculation that a larger dissolved gas concentration decreases the bubble lifetime by bringing an early rupture to the film. The next question arises naturally, that is, what is the essence of the particle? They could be solid particles due to impurities in the liquid, or they could be microbubbles. But one could easily tell that they are actually microbubbles mainly due to two reasons: First, all the particles observed are spherical, while solid impurities are often irregular in shape \citep{poulain2018ageing}. Second, upon film rupture, the particle suddenly ``disappears'' and no trace of impurities is observed in the film close to the hole (see Fig. \ref{fig:microbubble_rupture}). This suggests the gaseous nature of the particle. %It might seem counterintuitive at first sight for undersaturated liquid to have microbubbles in the bulk, but \citet{fang_formation_2018} has shown that stable bulk nanobubbles could exist in undersaturated liquids. We also observed microbubbles in the bulk of undersaturated (and also oversaturated) liquids, please see Supplementary Materials for details. 

\begin{figure}[h]
    \centering
    \includegraphics[width=0.4\textwidth]{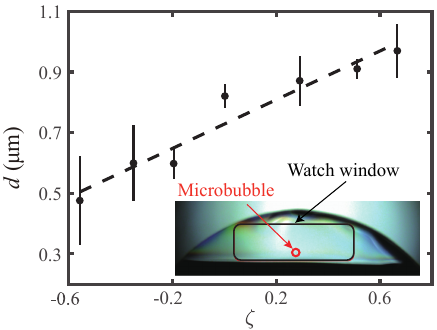}
    \caption{Microbubble diameter $d$ in the late stage of isopropanol bubble lifetime at different gas oversaturation $\zeta$. Each point represents the mean of 5 to 8 measurements and in total 52 microbubbles were measured. To make the diameter measurement accurate, the magnification was adjusted so that the bubble diameter occupies at least 10 pixels in the camera. The dashed line is a linear fit of the data points.}
    \label{fig:microbubble_stats}
\end{figure}

The diameters of the microbubbles $d$ are measured by monitoring the central region of the bubble, see Fig. \ref{fig:microbubble_stats}. The red box in the inset shows the monitored region. It can be seen that the average diameter of the microbubbles $\overline{d}$ increases with the dissolved gas concentration (represented by the gas oversaturation $\zeta$) almost linearly. This can be explained as follows. These microbubbles nucleate on the container wall, rise and grow in the bulk until they reach the surface. Some of them could enter the bubble cap and continue to rise. But given the very short duration when the microbubbles are in the bubble cap (\SI{2.4}{ms} for example, see Fig. \ref{fig:microbubble_rupture}), their size change in the bubble cap can be neglected. Thus, the microbubbles only grow in size when they are in the bulk. We then estimate the average diameter of the microbubble. Since the microbubbles nucleate on the container wall, it is reasonable to assume that their residence time in the bulk is the same. Further assuming that the growth of the microbubble is dictated by diffusion (given their small size and small rising velocity), we have $\overline{d}\sim c$ according to \citet{epstein1950stability}. 

\begin{figure}[H]
    \centering
    \includegraphics[width=0.5\textwidth]{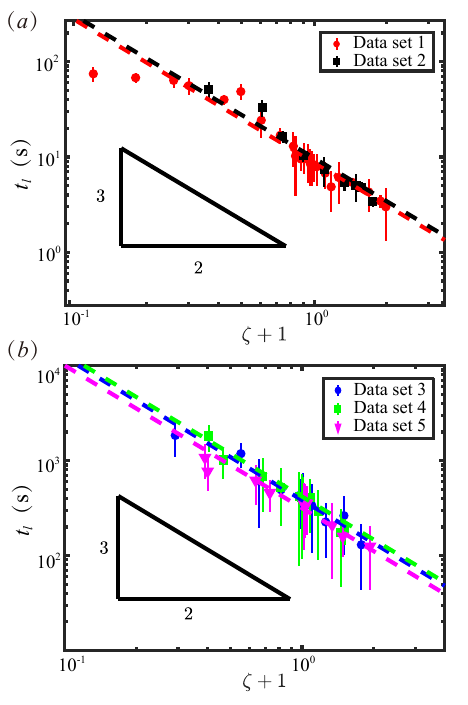}
    \caption{Dependence of bubble lifetime $t_l$ on dissolved gas concentration represented by the gas oversaturation $\zeta$. Data sets 1 and 2 in ($a$) are obtained from experiments in a petri dish with a plastic tube. Data set 1 is replotted from Fig.~\ref{fig:lifetime}. Data sets 3, 4 and 5 in ($b$) are from experiments in glass containers without the plastic tube. The detailed experimental conditions are shown in Table \ref{tab:parameters}. The dashed lines all have slopes of $-3/2$.\color{black}}
    \label{fig:lifetimeLogscale}
\end{figure}

Now we further analyze the dependence of bubble lifetime on dissolved gas concentration. The microbubbles must be large enough to trigger the rupture of the film. It is then reasonable to assume that the film will definitely rupture if the (average) microbubble diameter is larger than the film thickness by a certain factor, or in other words, $\overline{d}/h_r \approx k$ where $k>1$ is the factor. Actually, this can be confirmed by comparing the microbubble diameter shown in Fig. \ref{fig:microbubble_stats} with the film thickness at spontaneous rupture shown in Fig. \ref{fig:initial_thickness}($b$). For example, the average bubble diameter at $\zeta\approx0.45$ (or equivalently $c_\mathrm{o}\approx \SI{12}{mg/L}$) is $d\approx\SI{0.9}{\micro\meter}$, larger than the film thickness at rupture $h_r\approx\SI{0.7}{\micro\meter}$. Coming back to the analysis, $\overline{d}/h_r \approx k$ also means that $\overline{d}\sim h_r$. Because $\overline{d}\sim c$, we have $h_r \sim c$ (see Fig. \ref{fig:initial_thickness}($b$)). For volatile liquids, it has been found that the film thickness of a surface bubble follows $h\sim t^{-2/3}$ \citep{lhuissier2012bursting, poulain2018ageing}. So at the instant when the film ruptures, it follows $h_r\sim t_l^{-2/3}$. Substituting the dependence of film thickness at rupture $h_r$ on the dissolved gas concentration, we obtain $t_l\sim c^{-3/2}$. Since $c_\mathrm{sat}$ does not change (because the temperature and atmospheric pressure do not change during the experiments), we have $t_l\sim (\zeta+1)^{-3/2}$.The bubble lifetime data shown in Fig. \ref{fig:lifetime} is replotted in Fig. \ref{fig:lifetimeLogscale}($a$) as the red dots in log-log scale (data set 1). Another set of experiments were performed nine months later, as shown by the black squares (data set 2). It is found that this scaling law $t_l\sim (\zeta+1)^{-3/2}$ fits the two sets of experimental results well.

To check the influence of container type, liquid pool depth and bubble size  on the bubble lifetime, we performed 3 more groups of experiments in glass containers made of quartz, while changing pool depth and bubble size, see Fig. \ref{fig:lifetimeLogscale}($b$) for the experimental results (Data sets 3 to 5) and Table \ref{tab:parameters} for the detailed experimental conditions of all five data sets. First of all, despite all these changes, the bubble lifetime still follows the same scaling $t_l\sim (\zeta+1)^{-3/2}$. Therefore, this scaling is independent of the container type, pool depth and bubble size. But these factors influence the absolute lifetime of surface bubbles. First, notice that bubble lifetime in Fig. \ref{fig:lifetimeLogscale}($b$) is almost 40 times larger than that in Fig. \ref{fig:lifetimeLogscale}($a$). The primary change between \ref{fig:lifetimeLogscale}($a$) and ($b$) is that the container is changed to glass containers made of quartz, without using the plastic tube. This confirms that the microbubbles are mainly nucleated on the container wall (also on the plastic tube). The petri dish and the plastic tube are both made of polyethylene (PE), they are rougher than quartz, thus it is much easier for them to nucleate more and larger microbubbles than quartz, this leads to the drastic decrease in the bubble lifetime. Second, by comparing data sets 4 and 5, one finds that the bubble lifetime decreases while increasing the pool depth. This is easy to understand because larger pool depth means longer residence time of microbubbles, thus microbubbles are generally larger, leading to earlier rupture of surface bubbles. Finally, the lifetime of data set 4 (bubble radius $R\approx\SI{2.5}{mm}$) is slightly larger than that of data set 3 ($R\approx\SI{2}{mm}$), but the difference is too small to draw any conclusions. The fact that microbubbles mainly nucleate on the container wall also explains why for very long-lived bubbles ($t_l$ can reach $\approx\SI{1800}{s}$ in Data set 3), gas exchange with air of the liquid pool does not change the bubble lifetime: Because at $t=\SI{1800}{s}$, gas exchange with the quiescent air (mainly due to diffusion) only influences the gas concentration of a thin liquid layer of thickness $\delta\sim\sqrt{Dt}\sim\SI{1.3}{mm}$ on top of the liquid pool ($D$ is the diffusion coefficient of oxygen in isopropanol and is estimated as $\sim\SI{1e-9}{m^2/s}$). The liquid pool is \SI{20}{mm} in depth, so that the gas concentration of the liquid at the bottom -- where most of the microbubbles are nucleated -- does not change due to gas exchange with air. Notice that by doubling the pool depth to \SI{40}{mm}, the bubble lifetime decreases by a factor of \SI{25}{\%} (from $\approx\SI{417}{s}$ to $\approx\SI{316}{s}$ at $\zeta\approx0$). Thus this thin liquid layer $\delta\sim\SI{1.3}{mm}$ of different gas concentration does not really change the average diameter of microbubbles, which in turn does not really change the bubble lifetime. Therefore, the gas exchange with quiescent air has negligible effect on bubble lifetime. 
\color{black}

\begin{table*}[h]
\centering
\caption{Summary of experimental parameters for different data sets.}
\label{tab:parameters}
\small
\begin{tabular*}{\textwidth}{@{\extracolsep{\fill}} 
    c 
    >{\centering\arraybackslash}m{2.4cm}   % Liquid container 列，自动换行
    c 
    S[table-format=2.0] 
    c 
     >{\centering\arraybackslash}m{2cm} 
    c 
     >{\centering\arraybackslash}m{2cm} 
    c 
    >{\centering\arraybackslash}m{2.4cm}   % Acquisition time 列，固定 2.4cm，自动换行
    @{}}
\toprule
Data set & Liquid container & Material & {Pool depth} & {Bubble volume} & {Bubble radius} & {Repetitions} & Date of \\
No.      &                  &          & {(mm)}       & {(\si{\micro\liter})} & {(mm)}          & per data point& measurement \\
\midrule
1 & Petri dish with plastic tube         & PE     & 10 & 90  & $\approx$4   & 13-17 & Aug., 2024 \\\hline
2 & Petri dish with plastic tube         & PE     & 10 & 90  & $\approx$4   & 14-17 & May, 2025 \\\hline
3 & Glass container without plastic tube & Quartz & 20 & 25  & $\approx$2   & 15-17 & Nov., 2025 \\\hline
4 & Glass container without plastic tube & Quartz & 20 & 55  & $\approx$2.5 & 15-17 & Nov., 2025 \\\hline
5 & Glass container without plastic tube & Quartz & 40 & 55  & $\approx$2.5 & 16-19 & Nov., 2025 \\
\bottomrule
\end{tabular*}
\end{table*}

\section{Conclusions}
In summary, we have investigated how the surface bubble lifetime $t_l$ in volatile liquids depends on the dissolved gas concentration $c$. The initial film thickness of the bubble cap is found to be $h_i\approx\SI{0.93}{\micro\meter}$, independent of the dissolved gas concentration. Yet the film thickness at rupture increases with the dissolved gas concentration. High speed recordings reveal that microbubbles are responsible for the film rupture. Further, it is found that larger microbubbles at increased gas concentration trigger bubble bursting at earlier times, thus leading to a decrease in the bubble lifetime. Combined with the film thinning rate of surface bubbles in volatile liquids, we find that the dependence of bubble lifetime on dissolved gas concentration follows $t_l\sim c^{-3/2}$, which fits well with the experimental results. The scaling is also found to be independent of container type, pool depth and bubble size. The container type influences the absolute value of bubble lifetime drastically, confirming that the microbubbles mainly nucleate from the container walls. The pool depth also influences the bubble lifetime by changing the residence time of microbubbles: deeper pool leads to shorter bubble lifetime. The influence of bubble size on its lifetime is too small to draw any meaningful conclusions. Our findings may provide new insight on surface bubble lifetime and foam stability. Especially, carbonated beverages are often oversaturated, thus the extent of oversaturation may influence the stability of the foam.

\section{Data Availability}
The data supporting the findings of this study are available from the corresponding author upon reasonable request.

\section{Author Contributions}
\textbf{Xin Li}: Investigation, Data curation, Visualization, Writing original draft. \textbf{Yanshen Li}: Conceptualization, Methodology, Supervision, Writing – review and editing.

\section{Acknowledgments}
We acknowledge the financial support from the National Natural Science Foundation of China under grant No. 12272376. We also thank Li-Chen Huang for his help on preparing the figures. 

\appendix
\section{Different gas components should dissolve in or get released from the liquid at the same ratio}
\label{app:A}
According to Henry's law, we have
\begin{equation}
c_\mathrm{o}=p_\mathrm{o}H_\mathrm{o},
\end{equation}
where $H_\mathrm{o}$ is the Henry's law solubility constant for oxygen and $p_\mathrm{o}$ is the partial pressure of oxygen on top of the liquid.

Similarly, for nitrogen, we have 
\begin{equation}
c_{N}=p_{N}H_{N},
\end{equation}
where $c_{N}$ is the dissolved nitrogen concentration in the liquid, $H_{N}$ is the Henry's law solubility constant for nitrogen and $p_{N}$ is the partial pressure of oxygen. Then the ratio of dissolved nitrogen to oxygen is
\begin{equation}
\frac{c_N}{c_\mathrm{o}}=\frac{p_N}{p_\mathrm{o}}\frac{H_N}{H_\mathrm{o}}.
\end{equation}

Since the pool liquid is exposed to air in all the experiments, even during depressurization and pressurization, the pressure ratio $p_{N}/p_\mathrm{o}$ does not change. Henry's law solubility constant $H$ depends on temperature but the temperature is kept the same for each experimental run in our experiments. It is also known that the Henry's law solubility constant $H$ is independent on pressure when the pressure is not too large. Actually, we can estimate the change of $H_N/H_\mathrm{o}$ upon changing pressure using the Krichevsky–Kasarnovsky equation \citep{krichevsky1935thermodynamical, carrol1992system}. For each gas component $i$ ($i$ could be any gas component in the air) in isopropanol, the Krichevsky-Kasarnovsky equation states
\begin{equation}
\ln H_i(p) = \ln H_i^\infty + \frac{\bar{V}_i^\infty (p - p_l^\mathrm{v})}{RT},
\end{equation}
where $H_i(p)$ is the Henry’s constant for component $i$ at pressure $p$, $p_l^\mathrm{v}$ is the vapor pressure of isopropanol ($\sim \SI{0.058}{atm}$ at \SI{298}{K} according to \citet{majer1986enthalpies}), $H_i^\infty$ is the Henry’s constant at pressure $p_l^\mathrm{v}$, $\bar{V}_i^\infty$ is the partial molar volume of component $i$ at infinite dilution,  $R = \SI{82.057}{cm^3 \cdot atm \cdot mol^{-1} \cdot K^{-1}}$ and $T = \SI{298}{K}$. Then the ratio of Henry’s constants at pressure $p$ is 
\begin{equation}
\ln\frac{{H_{N}}(p)}{{H_\mathrm{o}}(p)}=\ln\frac{{H_{N}^\infty}}{{H_\mathrm{o}^\infty}}+\frac{\left (\bar{V}_N^\infty-\bar{V}_\mathrm{o}^\infty\right)\left(p-p_l^\mathrm{v}\right)}{RT}.
\label{eq:ratio_general}
\end{equation}
The difference of ratio $H_N/H_\mathrm{o}$ at two different pressures $p_1$ and $p_2$ is
\begin{equation}
\ln{\left ({\frac{H_N(p_2)/H_\mathrm{o}(p_2)}{H_N(p_1)/{H_\mathrm{o}}(p_1)}} \right )=\frac{(p_2-p_1)\left ({{{\bar{V}}_{N_2}^\infty}-{\bar{V}}_{O_2}^\infty}\right )}{RT}}
\label{eq:ln_diff}
\end{equation}
From \citet{battino1966solubility} we know that ${\bar{V}}_{N}^\infty = \SI{49.3}{cm^3/mol}$ and ${\bar{V}}_\mathrm{o}^\infty = \SI{42.5}{cm^3/mol}$. Let $p_2=\SI{4.5}{atm}$ (which is the largest pressure in our experiment) and $p_1=\SI{1}{atm}$, we obtain 
\begin{equation}
\ln\left ({\frac{H_N(p_2)/H_\mathrm{o}(p_2)}{H_N(p_1)/{H_\mathrm{o}}(p_1)}} \right )=0.00097.
\end{equation}
Therefore
\begin{equation}
{\frac{H_N(p_2)/H_\mathrm{o}(p_2)}{H_N(p_1)/{H_\mathrm{o}}(p_1)}}\approx1.00097.
\end{equation}
That is to say, by increasing the pressure from \SI{1}{atm} to \SI{4.5}{atm}, the ratio $H_N/H_\mathrm{o}$ only increases by less than \SI{0.1}{\%}, which is negligible. Thus, in our experiments, $H_N/H_\mathrm{o}$ can be considered as a constant. Consequently, $c_N/c_\mathrm{o}$ is a constant in our experiments. Naturally, for other gas components $i$ in the air, $c_i/c_\mathrm{o}$ is also a constant, then $c/c_\mathrm{o}$ is a constant, where $c=\sum c_i$ is the dissolved gas concentration which contains all gas components in the air. 

In conclusion, gas components in the air dissolve in or get released from the liquid at the same ratio, and the dissolved oxygen concentration $c_\mathrm{o}$ can represent the dissolved gas concentration $c$ in the liquid.

\color{black}
\bibliographystyle{elsarticle-harv}
\bibliography{References}

\end{document}